# 法律类案检索结果中性别偏置对用户认知的影响

张瑞喆 [1,2,3]　　艾清遥 [1,2,3]　　刘奕群 [1,2,3]　　吴玥悦 [1,2,3]　　王贝宁 [1,2,3]

(1. 清华大学 计算机科学与技术系, 北京 100084;

2. 清华大学 互联网司法研究院, 北京 100084;

3. 泉城实验室, 山东济南 250014)

**摘要:** 近年来, 法类案检索已经成为了法律从业者工作的重要的组成部分。法律类案检索的过程中, 搜索引擎从海量的数据中筛选出若干个相关案例, 提供给用户。然而, 这些案例是否具有性别偏置, 该种偏置是否会对用户认知产生影响, 却是一个未知数。作者设计了一个全新的用户实验框架, 模拟法官进行类案检索时阅读相关案例的过程。作者邀请了 72 位具有法律背景的志愿者进行实验。志愿者被要求模拟法官的身份就指定的 3 个案例进行类案检索, 并对案例中的被告人量刑。作者对任务和类案检索结果中被告人的性别均进行了编辑, 以统计类案检索结果中的性别偏置对被试认知产生的影响。实验结果显示, 类案检索结果中的性别偏置对法官的认知不产生显著的影响。

**关键词:** 法律类案检索; 用户实验; 公平性

**中图分类号:** TP391　　　　**文献标识码: A**

# Gender Biased Legal Case Retrieval System on Users' Decision Process

Ruizhe Zhang[1,2,3], Qingyao Ai[1,2,3], Yiqun Liu[1,2,3], Yueyue Wu[1,2,3], Beining Wang[1,2,3]

(1. Department of Computer Science and Technology, Tsinghua University, Beijing 100084, China;
2. Institute for Internet Judiciary, Tsinghua University, Beijing 100084, China;
3. Quan Cheng Laboratory, Jinan Shandong 250014, China)

**Abstract:** In the last decade, legal case search has become an important part of a legal practitioner's work. During legal case search, search engines retrieval a number of relevant cases from huge amounts of data and serve them to users. However, it is uncertain whether these cases are gender-biased and whether such bias has impact on user perceptions. We designed a new user experiment framework to simulate the judges' reading of relevant cases. 72 participants with backgrounds in legal affairs invited to conduct the experiment. Participants were asked to simulate the role of the judge in conducting a legal case search on 3 assigned cases and determine the sentences of the defendants in these cases. Gender of the defendants in both the task and relevant cases was edited to statistically measure the effect of gender bias in the legal case search results on participants' perceptions. The results showed that gender bias in the legal case search results did not have a significant effect on judges' perceptions.

近年来，随着法律类案检索技术的飞速发展和相关政策的落地，类案检索技术已经成为了法律从业者工作的重要组成部分。法律类案检索技术基于司法公开制度，利用搜索引擎等相关信息技术，为法律从业者的工作提供了便利。

在法律类案检索系统中，用户首先需要输入案件的案情描述，系统自动在海量的公开判决书中进行检索，并将和用户输入的案情描述最相关的数个案例提供给用户。法官、律师等从业人员可以通过检索，快速获取相关案例的推理过程和判决结果。法律类案检索的结果已经成为了从业人员高质量的参考材料，能帮助他们更快更好地完成工作。

在法律类案检索的过程中，搜索引擎会通过一定的方法检索出数个到数十个和用户查询的案情描述相关的判决文书。然而，这些文书往往是根据文本相关性和案由相关性，从海量的候选案例中筛选出来的。一个重要的问题是：在筛选的过程中，并没有考虑到筛选出的案例中，判决是否存在过轻或过重。设想下面的这种情况：对某被告人涉嫌盗窃罪的案情描述，法官在未受到裁判文书影响时，将对被告人判处有期徒刑两年。但在法律类案检索系统介入后，该法官浏览了 4 个和该案件相关的类案检索结果。四个案例的被告人性别和判决结果如下表所示。

表 1：检索结果示例
Table 1: Example of search results.

| 编号 | 1 | 2 | 3 | 4 |
|---|---|---|---|---|
| 性别 | 男 | 男 | 女 | 女 |
| 刑　期（月） | 30 | 32 | 20 | 16 |

那么，法官就有可能受到类案检索结果的影响，认为在盗窃罪案件中，男性应当被判决更长的刑期。从而对正在审理的案件中的男性被告人判处更长的刑期，对正在审理的案件中的女性被告人判处更短的刑期

如果这种情况是真实存在的，将会构成法律类案检索结果对司法裁判公正性的影响。因此，非常有必要进行相关的研究，以确定这种情况是否是真实存在的，用户受到的影响程度有多大。

按照我们的调研，这种情况是否存在，以及其对法官的影响，尚未被研究过。我们将在这篇文章里，对相关的问题进行研究。

我们发现，由于以下 5 个方面的原因，在真实的场景下进行相关的统计存在较大的困难。1. 商业法律类案检索引擎的用户群体复杂，通过其日志数据难以锁定我们需要研究的用户群体（专业的法律从业人员）。2. 商业的法律类案检索引擎中，用户可能会对案情描述进行反复的修改，多次进行查询，这些查询的结果对用户造成的可能的影响将叠加起来，为我们定量分析造成困难。3. 商业法律类案检索引擎是在线系统，直接介入并修改提供给用户的检索结果会影响用户体验，且如果该修改会对用户的认知产生影响，则还存在相关的伦理问题。4. 商业法律类案检索系统中，不同的用户几乎不可能提出同样的案情描述进行查询，因此提供给他们的结果也就不会完全一致。这就导致了我们对于每一个查询词，只能取得一个用户的检索行为数据。5. 商业法律类案检索系统只提供结果，而不收集用户对其案情描述所做出的判断，这就使得通过相关日志，无法得知检索结果是否对用户造成了影响，影响程度为何。

因此，我们设计并搭建了一个全新的法律类案检索实验系统，确保了上面的五点困难是在受控环境下的，进而开始研究。1. 该实验的被试从具有法律专业背景的人员中招募，以确保其行为符合专业人员的行为特征。2. 同时，系统支持预设任务，要求被试就我们给出的任务进行查询。3. 由于我们的案例是虚拟的，并非用户需要真实参与判决的案例，我们提供的结果中包含的任何偏置不会对用户的实际工作造成影响。4. 我们可以为所有的被试预设相同的案例，以保证单个案例下收集到的样本数量。5. 实验系统对多方面的数据进行了收集，包括判决刑期、问卷结果等，解决了用户反馈收集困难的问题。

在这个系统上，我们设置了 3 个不同的案情描述，并为每一个案情描述设置了 6 个类案检索结果。用户在阅读案情描述后，将依次查看类案检索结果。查看完毕后，用户被要求对预设的案情描述中的被告人给出判决刑期及理由。

在提供给不同用户的案情描述和类案检索结果中，被告人的性别及 6 个类案检索结果中的被告人的性别均经过我们人为的编辑。不同的被试

看到的案例中，被告人性别可能存在不同，其他内容完全相同。我们招募了 72 个被试并对他们分组，不同组用户阅读的案例的性别组合有所区别。

我们设计了 6 个研究问题（research question，RQ），并基于上述实验收集到的数据对这六个问题进行了统计分析。问题涵盖了被告人性别对判决结果的影响、类案检索结果中的性别偏置对判决结果的影响、被试性别对判决结果的影响、被试和被告性别的组合对判决结果的影响以及被试是否意识到了我们给出的案例有偏等 5 个方面。

研究结果表明，被试阅读类案检索系统提供的检索结果后，对实验案例中的被告人给出的判决结果（刑期）在各种设定下的差异均不存在显著性。因此，类案检索系统的研发人员无需对这方面过分担忧，即使提供的检索结果中存在性别偏置，也不会影响专业法律从业用户在检索后的判断。

我们的贡献主要体现在四个方面： 1. 注意到了法律检索系统对用户认知产生影响，进而影响到司法公正的一种可能性。 2. 设计并实现了一个实验框架，模拟法律类案检索的过程。 3. 招募被试进行实验，形成了第一个法律类案检索领域中考虑被告人性别和类案检索结果中被告人性别的数据集。 4. 对数据集进行了全方位的分析，给出了类案检索系统的研发人员无需考虑性别偏置对用户的影响的重要结论。

# 1 相关工作

在信息检索领域，前人已经进行过许多的探索和研究。在上世纪末，研究人员采用 BM25[1, 2]，TF-IDF[3]等基于短语匹配（term-level）的方法进行对文献进行索引。本世纪以来，许多学习排序（Learning to Rank，LTR）的方法[5,6]，用于信息检索。近十年来，神经网络（Neural Network）也被应用于信息检索领域，卷积神经网络[7]（CNN），循环神经网络（Recurrent neural network，RNN）[8,9]等模型也被用于信息检索领域。近几年，一些基于自然语言的方法（NLP model），如基于变换器的双向编码器表示技术（Bidirectional Encoder Representations from Transformers， BERT）及其变形[10,11]也被用于信息的编码和索引。这些模型在通用的信息检索中发挥了良好的作用，然而，这些模型对法律检索的考虑是不足的，没有考虑到法律类案检索场景下用户的信息需求。

在法律类案检索的场景下，科研人员也进行了不懈的探索。早在 1987 年，将检索技术应用于法律领域就受到了学界和业界的关注[12]。Turtle 等人[13]提出将文本检索的方案应用于法律领域。Van 等人[14]在 2017 年提出了法律领域的相关性的定义。一些基于知识和逻辑的方法也被提出并用于法律检索[15,16]。同时，考虑了文本长度特点的检索方法[17]、综合考虑文本相似性和案由相似性的方法[18]。这些模型主要考虑了如何为用户提供契合其信息需求的结果，但忽略了法律领域中最重要的问题——公平。检索结果是否有信息偏置，这种信息偏置是否会导致判决结果的公平性存在疑点，这是司法领域非常重要的一个问题。只有妥善解决这个问题，法律类案检索才能更好地服务于法律从业人员。

研究人员也在检索公平性上做过相关的研究。检索公平性聚焦于检索结果是否得到了公平的展示机会。公平性的概念最早来自于分类算法[19]。Geyik 等人的工作中，提出了检索领域的公平性的概念[20]。Fu 等人利用知识图谱的方法优化检索结果的公平性[21]。但这些研究都没有考虑到法律文书的特性，许多重要的信息（如罪名、刑期）没有得到应有的重视。

在判决公平性领域，研究则主要关注性别、年龄、地区等因素是否会对判决结果产生影响。研究者们对司法判决中的性别偏置进行了充分的研究，从多个不同的角度诠释了被告人性别对判决结果可能的影响[22,23,24]。另一些研究对法官的性别和法院的地区对判决结果造成的影响进行了研究和分析[25,26]。其他一些研究关注了性别、年龄和年龄对于判决结果的共同作用[27]。这些文献对相关因素如何影响判决结果做出了论述，但据我们所知，还没有相关的文献对类案检索结果是否会影响公平性做出判断和验证。

# 2 实验框架设计

## 2.1 实验流程

我们设计了一个全新的实验框架，模拟司法裁判人员在工作中进行类案检索时的情形。

具体来说，被试需要先用浏览器登录平台，然后开始实验的主体部分。实验的主体部分结束

后，被试还需要完成一份调查问卷，以帮助我们了解其认知和完成情况。

实验的主体部分由若干个任务构成，任务依次进行，完成了上一任务才能开启下一任务。每个任务的内容均为要求被试阅读一段案情描述，并给出判决结果、判决考量的因素并简述判决理由。每个任务的时间限制（上限和下限）由管理员设定。

为模拟用户通过法律类案检索等手段完成任务的过程，用户在了解完任务内容后，我们会为其提供 6 个参考案例，及检索法律所用的网站。我们的实验网站截图如图 1。

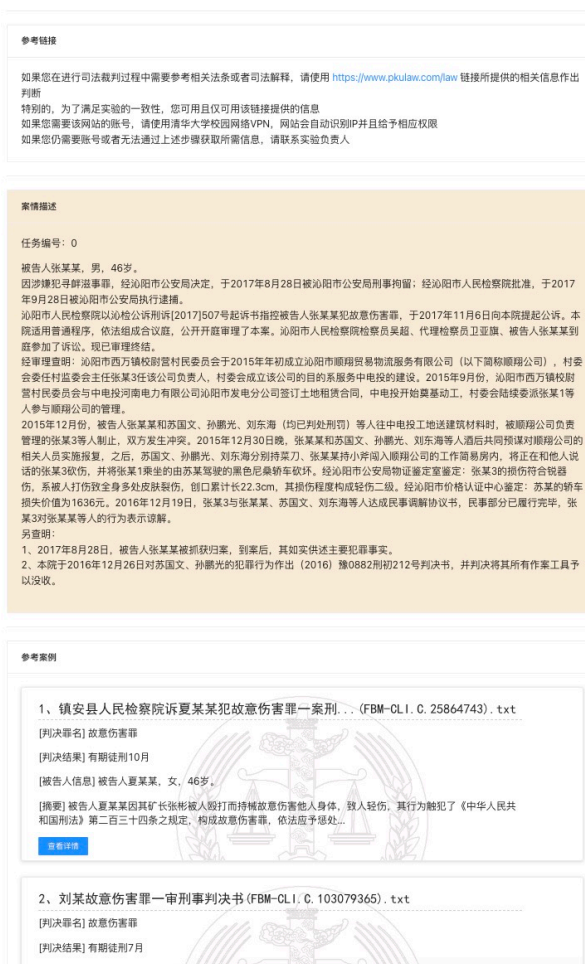

图 1：任务、链接、类案展示页面

Figure 1: Page displaying tasks, links, and case categories.

在任务完成时，被试需要提交其关于这个任务给出的量刑结果，实验网站截图如图 2。

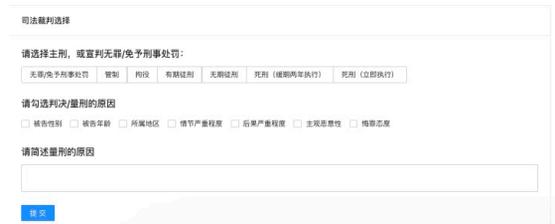

图 2：结果提交模块

Figure 2: Module for submitting results.

## 2.2 实验平台

为了便于大规模开展试验，我们设计并开发了一个实验平台。该平台使用前后端分离的设计方法，前端使用 Vue 搭建，后端使用 Django。

用户可以在实验上输入实验账号并登录进行实验。用户阅读实验任务和阅读类案检索结果的步骤在平台上完成，查询法条的步骤跳转到外部完成。每个任务结束时，被试需要在平台上给出任务中被告人的量刑，勾选考虑的因素，并简述理由。实验结束后，用户还需完成一份调查问卷。

上述所有的实验数据均被自动收集和记录，此外我们还记录了用户完成每个任务所消耗的时间，对于完成时间过短的用户，平台对其进行提示，要求其认真完成实验。

## 3 实验内容

### 3.1 任务设置

在本实验中，我们需要选取适当数量和内容的任务案例，以使得实验时长合理。

在对于实验时长设置的方面，经过预实验和充分的评估，我们认定一位具有法律知识背景的被试完成单个任务需要花费大约 25-35 分钟时间。该时间包括了被试理解任务、查询法条和司法解释、阅读 6 个类案检索结果的时间。因此，设置 3 个实验任务，以将实验时间控制在 2 小时以内，是合理的设计。

在案由涉及的案件方面，综合考虑了诉请复杂程度、裁判结果形式、司法裁判过程、用户阅读困难程度等方面的因素。我们选取 3 个真实的刑事案例作为检索任务。每个案例分别涉及到一名被告人和一项犯罪。并且，这三个案例的法定量刑区间均应于有期徒刑一个月至有期徒刑二十年之间。这样我们可以对被试给出的刑期进行数学上的定量分析。在案由方面，我们综合考虑了暴力犯罪和经济犯罪，分别选取了 2 个和 1 个案例作为任务。所有的任务均来自于 LeCard[28] 中真实的司法裁判文书的"案情描述"部分，表

2 种列举了这些案例的基本信息。

表 2：任务案例基本情况
Table 2: Basic information of task cases.

|  | 案由 | 原判决 | 案情描述长度（字） | 地区 | 审理程序 |
|---|---|---|---|---|---|
| 案例 1 | 故意伤害 | 有期徒刑 12 月 | 886 | 河南沁阳 | 一审 |
| 案例 2 | 放火 | 有期徒刑 36 月 | 772 | 山东高密 | 一审 |
| 案例 3 | 职务侵占 | 有期徒刑 120 月 | 1004 | 湖南岳阳 | 一审 |

### 3.2 检索结果设置

在检索结果设置方面，我们需要就每个任务设置相应的类案检索结果。我们按照以下规则从 LeCard 对该任务提出的相关判决书中选择。

1. 选择单人、单罪的判决书。
2. 选择案由与任务相一致的判决书。
3. 选择法定量刑区间与任务一致的判决书。
4. 选择案情简单，文本长度较短（不超过 3000 字）的判决书，以减轻被试阅读压力。
5. 选择文本内容不会暴露被告人性别、年龄等信息的判决书（例如，提及"被告人的妻子"会暴露被告人是男性）。

我们从符合以上 5 条规则的，选择相关性最高的 6 份判决书（根据 LeCard 中的标注）。

由于司法公开制度相关的隐私政策，这些判决书中已经隐去被告人的年龄、性别。在本实验中，我们为所有用户提供相同的类案检索结果，但为判决书和任务中的被告人均根据我们的需求加上性别。具体来说，我们的实验设定有不同的 12 种，其构成如表 3 所示。其中，量刑较轻和量刑较重分别指 6 份判决书中刑期较短的 3 份和刑期较长的 3 份。表格中的数字表示的是提供给被试的案例中，3 份判决书中的性别分布。

每位被试会被随机分配到一组实验设定，同一个被试的 3 个任务的实验设定相同。

表 3：实验设定分布表
Table 3: Distribution of experimental settings.

|  |  |  | 量刑较轻 |  | 量刑较重 |  |
|---|---|---|---|---|---|---|
| 序号 | 任务中的被告性别 | 特征 | 男 | 女 | 男 | 女 |
| 1 | 男 | 标准 | 1 | 2 | 1 | 2 |
| 2 | 男 | 标准 | 2 | 1 | 2 | 1 |
| 3 | 男 | 男性轻判 | 3 |  |  | 3 |
| 4 | 男 | 男性偏轻 | 2 | 1 | 1 | 2 |
| 5 | 男 | 女性轻判 |  | 3 | 3 |  |
| 6 | 男 | 女性偏轻 | 1 | 2 | 2 | 1 |
| 7 | 女 | 标准 | 1 | 2 | 1 | 2 |
| 8 | 女 | 标准 | 2 | 1 | 2 | 1 |
| 9 | 女 | 男性轻判 | 3 |  |  | 3 |
| 10 | 女 | 男性偏轻 | 2 | 1 | 1 | 2 |
| 11 | 女 | 女性轻判 |  | 3 | 3 |  |
| 12 | 女 | 女性偏轻 | 1 | 2 | 2 | 1 |

### 3.3 被试

在招募被试方面，考虑到司法裁判任务的特殊性，被试应当具备相应的法律知识背景，方能胜任该任务。我们在全国范围内，公开招募了法律专业的被试二百余人，其中研究方向为刑法学的 92 人。

我们从 92 位具有刑法学背景的报名人员中，

选择了 72 位作为本实验的被试。在实验前，我们向每位被试发放了在线实验系统的用户账号，被试利用个人电脑登录实验账号远程进行实验。

实验完成后，我们为每位被试提供了人民币 200 元的报酬。

### 3.4 实验后问卷

为全方位了解被试在检索过程中的情况，确

保形成一份完整的数据集，我们在实验结束后，要求被试完成一份带有 8 个问题的问卷。每个问题的形式均为："你认为应当如何评价＿＿"。选项包括为 5 个等级，包括"完全不同意"，"有一点不同意"，"一般"，"有一点同意"和"完全同意"。

8 个问题考虑了多个不同的方面，具体内容为：

1. 对于整个任务过程，你感到满意。
2. 对于做出的判决结果，你感到有信心。
3. 完成任务的过程中消耗了大量的脑力（如：计算、思考、决定、记忆等）。
4. 任务给出的参考案例是公平的。
5. 在给出参考案例时，系统提供的案例更倾向于对男性（或女性）量刑更重。
6. 在给出参考案例时，系统提供的案例更倾向于对年轻人（或年长者）量刑更重。
7. 在给出参考案例时，系统提供的案例更倾向于对某些地区的被告人量刑更重。
8. 参考案例协助你完成了任务。

该评价是对整个实验的 3 个任务进行的总体评价，其中问题 5-7 用于下一章节中的 RQ5 和 RQ6。

# 4 研究问题及结果

本节中，我们将介绍我们从 2 方面提出的 6 个研究问题及它们的结果。

我们的研究问题（Research Question, RQ）包括以下 6 个。RQ1-RQ4 从不同维度考虑了可能存在的性别偏置对检索结果的影响。包括：

RQ1：被告人性别对刑期是否有影响？

RQ2：法律类案检索结果中的性别偏置和任务案例中被告人性别的联合分布对刑期是否有影响？

RQ3：被试性别对刑期是否有影响？

RQ4：被试性别和被告人性别的联合分布对刑期是否有影响？

同时，RQ5 和 RQ6 对来自于用户问卷反馈的信息进行了分析。以解答用户是否能意识到自己阅读了有偏的信息。

RQ5：相比于阅读了无偏的类案检索结果的被试，阅读了有偏组别的被试是否更倾向于类案检索结果带有性别偏置？

RQ6：用户是否更倾向于认为，系统给出的类案检索结果带有性别偏置而不是地区或年龄偏置？

## 4.1 RQ1：被告人性别对刑期是否有影响？

在司法领域，根据法律面前人人平等的精神，被告人的性别不应当对有期徒刑的刑期产生影响。然而，在司法实践中，这种影响又未必对我们所愿并不存在。一些研究认为，在某些条件下被告人性别对刑期有影响[25,26]。另外一些研究认为，被告人性别对刑期没有影响。

我们对从 72 位被试提供的刑期数据进行了分析，以验证被告人性别是否会对刑期产生影响。在 72 位被试中，有 36 位被试的三个查询任务中的被告人（被设置）为女性（下文简称：女性被告），另外 36 位被试查询任务中的被告人（被设置）为男性（下文简称：男性被告）。

我们对这些被试提出给出的刑期进行了分组分析，统计了三个任务中男性被告被判处的平均刑期和女性被告被判处的平均刑期及它们的标准差，结果如表 4 所示。

表 4：不同性别的被告被判处的平均刑期及其标准差（单位：月）

Table 4: Average sentence length and standard deviation for defendants of different genders (in months).

| | | 男性被告 | | 女性被告 | |
| | | 平均刑期 | 标准差 | 平均刑期 | 标准差 |
|---|---|---|---|---|---|
| 案例 1 | 故意伤害 | 11.74 | 2.47 | 12.97 | 2.30 |
| 案例 2 | 放火侵占 | 39.78 | 7.39 | 37.00 | 11.16 |
| 案例 3 | 职务侵占 | 122.05 | 25.08 | 118.08 | 23.77 |

结果显示，在故意伤害罪的案例中，女性被判处的刑期相对较长。而在放火罪和职务侵占罪的案例中，情况则有所不同，男性被判处的刑期相对较长。注意到被试给出的刑期的标准差实际上达到了平均刑期的 17%~30%，而性别之间平均刑期的差距只有 3%~11%，这可能是由于随机分布的误差所导致的。

我们对三个任务的结果进行了 t 检验（t-test），其中原假设为"不同性别的被告人将被判处相同的刑期"，p 值在 0.16-0.25 之间，原假设均不能被拒绝。即，没有证据表明，被告人的性别会影响被试（法官）对其的量刑。

### 4.2 RQ2：法律类案检索结果中的性别偏置和任务案例中被告人性别的联合分布对刑期是否有影响？

进一步地，我们还探究了法律类案检索的结果中的性别偏置和任务案例中被告人的性别偏置的联合分布是否会对判决结果产生影响。

在我们的实验收集到的数据中，被试根据（男性轻判，无偏，男性重判）×（任务中被告为男性，任务中被告为女性）的联合分布分为了 6 组。每组人数均在 11-13 人之间。

我们统计了每组被试对三个任务案例分别判处的平均刑期，如表 5 所示。

表 5：根据（男性轻判,无偏，男性重判）×（任务中被告为男性，任务中被告为女性）联合分组后，每组的平均刑期（单位：月）

Table 5: Average sentence length (in months) for each group after combining based on (lenient sentencing for males, unbiased, harsh sentencing for males) × (defendant being male in the task, defendant being female in the task).

| 组别 | 男性轻判 | | 无偏 | | 男性重判 | |
| --- | --- | --- | --- | --- | --- | --- |
| 罪名 | 男性刑期 | 女性刑期 | 男性刑期 | 女性刑期 | 男性刑期 | 女性刑期 |
| 故意伤害 | 12.50 | 11.81 | 12.18 | 13.27 | 11.00 | 12.50 |
| 放火 | 36.33 | 37.00 | 42.36 | 37.25 | 40.00 | 39.83 |
| 职务侵占 | 128.18 | 117.91 | 122.18 | 117.08 | 116.33 | 119.15 |

我们使用 t 检验进行了以下两个方面的显著性检验。 1. 对于阅读每一类偏置结果的用户，其对任务案例中的男性被告人或任务案例中的女性被告人判处的刑期是否差异化？ 2. 相比于阅读了无偏的类案检索结果的用户，阅读了带有某种偏置的类案检索结果的用户是否会对任务案例中某一性别的被告人判处更长或更短的刑期？

t 检验结果显示，在这两个方面，均不存在统计意义上的显著差异。即：法律类案检索结果中的性别偏置和任务案例中被告人性别的联合分布对刑期没有影响。

### 4.3 RQ3：被试性别对刑期是否有影响？

接着，我们研究了被试的性别对刑期是否有影响。以往的一些研究中，研究人员发现被试性别可能在某些情况下影响到判决的结果[27]。

在本实验中，男性被试有 25 人，女性被试有 47 人。我们统计了男性被试对被告判处的平均刑期和女性被试对被告判处的平均刑期。结果如表 6 所示。与前面的结果相同，t 检验的结果表明差异不具有显著性。

表 6：被试给出的平均刑期（单位：月）

Table 6: Average sentence length provided by the participants (in months).

| 罪名 | 男性被试 | 女性被试 |
| --- | --- | --- |
| 故意伤害 | 11.81 | 12.28 |
| 放火 | 38.04 | 39.26 |
| 职务侵占 | 123.75 | 118.06 |

### 4.4 RQ4：被试性别和被告人性别的联合分布对刑期是否有影响？

在关于刑期是否会受到性别影响的最后一个 RQ 中，我们探究了被试性别和检索任务被告人的性别的联合分布对刑期是否有影响。

一个可能的猜想是，由于被试必然对两种性别有一定的社会印象。被试可能会出于对于某一性别的同情或厌恶，对这一性别的被告人重判/轻判。

在 25 名男性被试中，12 名的任务中，被告人性别为男性，其余 13 名任务中的被告人性别为女性。在 47 名女性被试中，22 名任务中被告人性别为男性，其余 25 名任务中的被告人为女性。

我们对三个任务分别统计了（被试男性，被试女性）×（被告男性，被告女性）的联合分布

下的平均刑期。结果如表 7 所示。在故意伤害任务中，男性被试对男性被告定的刑期较轻，在放火任务中，男性被试对女性被告量刑较轻，而在职务侵占案例中，女性被试对女性被告量刑较轻。

对于每一个任务中的 4 组变量，我们对其两两进行了 t 检验。t 检验的结果显示，任意两组结果变量的对比结果没有统计学意义上的显著性（p>0.05）。

表 7：（被试男性，被试女性）×（被告男性，被告女性）的联合分布下的平均刑期

（单位：月）

Table 7: Average sentence length (in months) under the joint distribution of (male participants, female participants) × (male defendants, female defendants).

|  | 故意伤害 | | 放火 | | 职务侵占 | |
|---|---|---|---|---|---|---|
|  | 男性被试 | 女性被试 | 男性被试 | 女性被试 | 男性被试 | 女性被试 |
| 男性被告 | 10.90 | 12.23 | 38.36 | 40.52 | 126.83 | 119.45 |
| 女性被告 | 12.58 | 12.33 | 37.75 | 38.16 | 120.66 | 116.84 |

### 4.5 RQ5：相比于阅读了无偏的类案检索结果的被试，阅读了有偏组别的被试是否更倾向于类案检索结果带有性别偏置？

接下来，我们从问卷的角度出发，探究阅读了带有偏置的案例的被试，是否倾向于认为类案检索结果带有性别偏置。

72 名被试中，32 名被试被分配到了有偏的类案检索结果，16 名被试被分配到了略有偏的类案检索结果，另外 24 名被试分配到了无偏的结果。

实验结束后的问卷要求被试对于类案检索结果是否有偏做出评价，我们将从无偏到有偏的 5 个等级用-2 到 2 中的 5 个整数表示。数字越小表示越无偏，数字越大表示越有偏。

每组的平均统计结果如表 8 所示。可以看出，对于无偏组，被试基本都认为结果是无偏的。而对于有偏和略有偏两组，相较于无偏组，总体上来说被试注意到了这种偏置，但从数值上而言可能他们对系统给出的案例是否有偏并不那么确信。

表 8：不同组别的用户是否感受到了偏置

Table 8: Whether users in different groups perceived bias.

| 实际偏置 | 无偏 | 略有偏 | 有偏 |
|---|---|---|---|
| 平均评分 | -1.34 | -0.68 | -0.65 |
| 标准差 | 1.04 | 1.15 | 1.16 |

显著性检验的结果表明，相较于无偏组，略有偏组和有偏组中被试对偏置的感受更加明显是具有统计学意义的（p<0.05）。但对于略有偏和有偏组之间的比较，并不能证明更强的偏置更容易被被试发现。

### 4.6 RQ6：用户是否更倾向于认为，系统给出的

类案检索结果带有性别偏置而不是地区或年龄偏置？

进一步地，我们利用问卷中的混淆问题来分析被试是否清楚地意识到了类案检索结果的偏置。这两个问题是：类案检索结果是否有年龄/地区偏置。事实上，在所有被试阅读的类案检索结果和任务中，被告人的年龄都被设定为 40-50 岁之间的随机数。而地区则使用了原判决书的地区，未进行编辑。其中，年龄偏置是指类案检索结果倾向于对某一（或某些）年龄段的

图 3 展示了问卷中"你认为应当如何评价：类案检索结果带有＿＿方面的偏置。"的统计结果。可以看出，相较于确实带有偏置的性别，用户也认为类案检索结果带有年龄、地区的偏置。并且，认为带有年龄和地区偏置的人数及其分布都和认为代表性别偏置的人数相当。

考虑和 RQ5 的结论综合来看，这说明被试不能准确地意识到偏置来源于何方面。

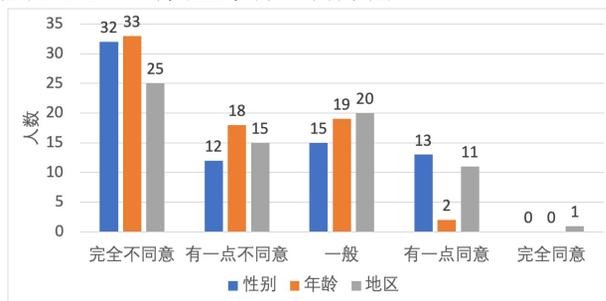

图 3：被试对每种偏置是否存在的同意程度

Figure 3: Participants' agreement level on the presence of each bias.

### 4.7 总结

综上所述，我们可以总结出两点结论。第一，各项结果均表明，不论法律类案检索结果向用户提供的检索结果是否存在性别偏置，均不会导致判决结果出现统计意义上的变化。第二，用户间

卷表明, 用户一定程度上能发现类案检索系统给出的结果是具有偏置的, 但是不能准确判断这种偏置的严重程度为何, 也不能准确判断出偏置主要是哪一个维度上的。

## 5 总结与未来工作

在本文中, 我们注意到法律类案检索结果中的性别偏置可能对判决结果产生影响, 妨碍司法公平公正。我们设计了一个用户实验框架, 并选择了 3 个不同案由的任务案例。我们招募了 72 名被试, 邀请他们在这个框架下完成这些任务和一份问卷。被试实验的过程被记录下来, 成为这方面的第一个数据集合。我们从 2 个方面提出了 6 个 RQ, 并通过统计结果回答了这些问题, 全面判断法律类案检索结果中的性别偏置是否会对判决公正产生影响。

此外, 我们的实验设定也存在一些局限性。包括被试是从法学院的学生中招募的, 而非真实的法官, 被试的性别并不均衡, 每个案例的判决时间只有半小时等。但我们认为这些设定是当前作为一个具有创新性的尝试的最好选择。这并不影响我们在本工作中作出的贡献。

如果未来有更多的机会, 我们将考虑招募更多的, 具有不同法律背景的被试, 进行更长时间的实验, 以分析系统对用户的影响。

## 参考文献

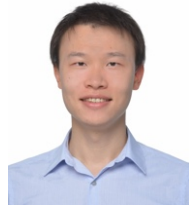
艾清遥（1992—），博士，助理教授，CCF 会员，主要研究领域为排序优化，信息检索等。
E-mail: aiqy@tsinghua.edu.cn

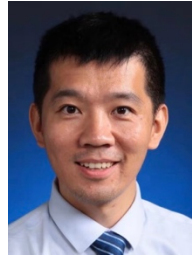
刘奕群（1981—），通信作者，博士，教授，CCF 会员，主要研究领域为信息检索、自然语言处理等。
E-mail: yiqunliu@tsinghua.edu.cn

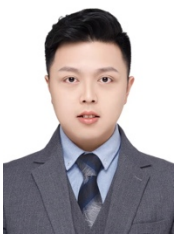
张瑞喆（1997—），博士研究生，主要研究领域为法律类案检索。
E-mail: u@thusaac.com